\documentclass[twocolumn,floats,showpacs,prl,amsmath,amssymb,floatfix,nofootinbib,balancelastpage]{revtex4}
\usepackage{psfig}

\newcommand{\ApJL}{Astrophys. J. Lett.}

\newcommand{\ApJ}{Astrophys. J.}
\newcommand{\PRL}{Phys. Rev. Lett.}
\newcommand{\PRD}{Phys. Rev. D}

\newcommand{\MNRAS}{Mon. Not. Roy. Astron. Soc.}

                              \newlength{\strikewidth}
                              \newlength{\strikelength}
\begin{document}

\title{Charged-particle decay and suppression of small-scale power}

\author{Kris Sigurdson}
\email{ksigurds@tapir.caltech.edu}
\affiliation{California Institute of Technology, Mail Code 130-33, Pasadena, CA
91125}
\author{Marc Kamionkowski}
\email{kamion@tapir.caltech.edu}
\affiliation{California Institute of Technology, Mail Code 130-33, Pasadena, CA
91125}


\begin{abstract}

We study the suppression of the small-scale power spectrum due
to the decay of charged matter to dark matter prior to recombination.
Prior to decay, the charged particles couple to the
photon-baryon fluid and participate in its acoustic
oscillations.  However, after decaying to neutral dark matter
the photon-baryon fluid is coupled only gravitationally to the
newly-created dark matter.  This generically leads to
suppression of power on length scales that enter the horizon
prior to decay.  For decay times of $\sim$$3.5$ years this leads
to suppression of power on subgalactic scales, bringing the
observed number of Galactic substructures in line with
observation.  Decay times of a few years are possible if the dark
matter is purely gravitationally interacting, such as the
gravitino in supersymmetric models or a massive Kaluza-Klein
graviton in models with universal extra dimensions. 
\end{abstract}


\pacs{98.80.Es,98.80.Cq,95.35.+d,98.62.Gq}

\maketitle

The standard inflation-inspired cosmological model, with its nearly scale-invariant power spectrum of primordial perturbations, is in remarkable agreement with observation.  It predicts correctly the detailed pattern of temperature anisotropies in the cosmic microwave background (CMB) \cite{CMB}, and accurately describes the large scale clustering of matter in the Universe \cite{LSS}.  However, on subgalactic scales there are possible problems with the standard cosmology that warrant further investigation.  Namely, the model overpredicts the number of subgalactic halos by an order of magnitude compared to the 11 observed dwarf satellite galaxies of the Milky Way \cite{excessCDMpower}.  Several possible resolutions have been proposed to this apparent discrepancy, ranging from astrophysical mechanisms that suppress dwarf-galaxy formation in subgalactic halos (see, for example, Ref.~\cite{AstroSol}) to features in the inflaton potential that suppress small-scale power  and thus reduce the predicted number of subgalactic halos \cite{Kamion2000}.

In this \emph{Letter}, we show that if dark matter is produced
by the out-of-equilibrium decay of a long-lived charged
particle, then power will be suppressed on scales smaller than
the horizon at the decay epoch.  Unlike some other recent
proposals, which suppress small-scale power by modifying the
dark-matter particle properties \cite{OtherMods}, ours modifies
the dark-matter production mechanism.
In the model we discuss here, prior to
decay, the charged particles couple electromagnetically to the
primordial plasma and participate in its acoustic oscillations.
After decay, the photon-baryon fluid is coupled only
gravitationally to the neutral dark matter.  This generically
leads to suppression of power for scales that enter the horizon
prior to decay. This suppression, reduces the amount of halo
substructure on galactic scales while
preserving the successes of the standard hierarchical-clustering
paradigm on larger scales.  Apart from the changes to the model
due to the decay process, we adopt the standard flat-geometry
$\Lambda$CDM cosmological model with present-day dark-matter
density (in units of the critical density) $\Omega_{d}=0.25$,
baryon density $\Omega_{b}=0.05$, cosmological constant
$\Omega_{\Lambda}=0.70$, Hubble parameter $H_{0}=72~{\rm
km\,s^{-1} Mpc^{-1}}$, and spectral index $n=1$.

\begin{figure}
\centerline{\psfig{file=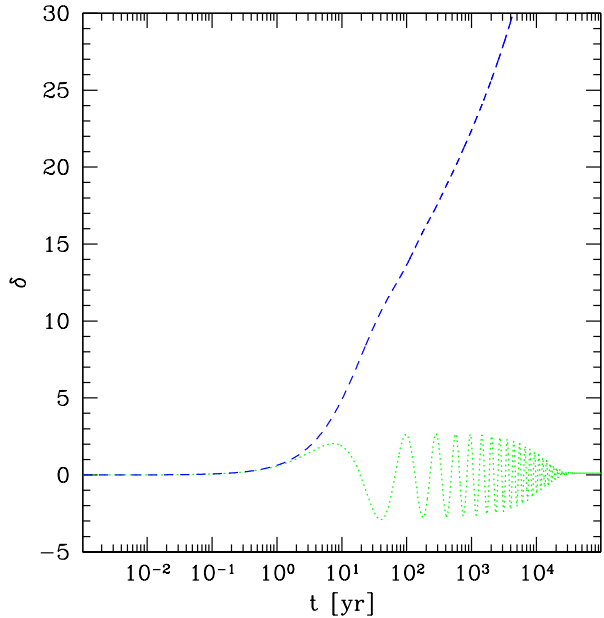,width=3.6285in,angle=0}}
\caption{The evolution of the comoving wavenumber $k=3.0~{\rm Mpc}^{-1}$ density perturbations in the early Universe for dark matter (dashed line) and baryons (dotted line).  The dark-matter perturbation always grows under the influence of gravity while the baryonic perturbation oscillates due to a competition between gravity and the photon pressure.}
\label{fig:delta}
\end{figure}

In the standard $\Lambda$CDM model the initial curvature perturbations of the Universe, presumably produced by inflation or some inflation-like mechanism, are adiabatic (perturbations in the total density but not the relative density between species) and Gaussian with a nearly scale-invariant spectrum of amplitudes.  These initial perturbations grow and react under the influence of gravity and other forces, with the exact nature of their behavior dependent upon the species in question.  Because dark-matter particles are, by assumption, cold and collisionless the fractional dark-matter-density perturbation $\delta_{d} \equiv \delta\rho_{d}/\rho_{d}$ can only grow under the influence of gravity.  The baryonic species, being charged, are tightly coupled by Coulomb scattering to the electrons, which are themselves tightly coupled to the photons via Thomson scattering.  The baryons and photons can thus be described at early times as a single baryon-photon fluid, with the photons providing most of the pressure and inertia and the baryons providing only inertia.  Gravity will tend to compress this baryon-photon fluid, while the radiation pressure will support it against this compression.  The result is acoustic oscillations, and the baryon density perturbation $\delta_{b} \equiv \delta\rho_{b}/\rho_{b}$ and photon density perturbation $\delta_{\gamma} \equiv \delta\rho_{\gamma}/\rho_{\gamma}$ will oscillate in time for length scales inside the horizon (on length scales larger than the horizon the pressure can have no effect). At early times these perturbations are very small and linear perturbation theory can be applied. This allows an arbitrary density field to be decomposed into a set of independently evolving Fourier modes, labeled by a wavenumber $k$.  Fig.~\ref{fig:delta} shows the growth of dark-matter perturbations under the influence of gravity, and the oscillatory behavior of the baryon perturbation for the same wavenumber. 

We choose to work in the synchronous gauge where the time slicing is fixed to surfaces of constant proper time so that particle decays proceed everywhere at the same rate.  In the synchronous gauge the standard linearized evolution equations for perturbations in Fourier space are (e.g., \cite{Ma95})

\begin{align}
\dot{\delta}_{d}  = - \theta_{d}-\frac{1}{2}\dot{h} \, ,
\quad
\dot{\theta}_{d} = -\frac{\dot{a}}{a}\theta_{d} \, ,
\label{eqn:dark_delta}
\end{align}
\begin{align}
\dot{\delta}_{b}  = - \theta_{b}-\frac{1}{2}\dot{h} \, ,
\end{align}
\begin{align}
\dot{\theta}_{b} = -\frac{\dot{a}}{a}\theta_{b} &+ c_{s}^2k^2\delta_{b} + \frac{4\rho_{\gamma}}{3\rho_{b}} a n_{e} \sigma_{T} (\theta_{\gamma}-\theta_{b}) \, ,
\label{eqn:baryon_theta}
\end{align} 
\begin{align}
\dot{\delta}_{\gamma} = -\frac{4}{3}\theta_{\gamma}-\frac{2}{3}\dot{h} \, ,
\end{align}
and
\begin{align}
\dot{\theta}_{\gamma} = k^2 \left( \frac{1}{4}\delta_{\gamma} - \Theta_{\gamma} \right) + a n_{e}\sigma_{T}(\theta_{b}-\theta_{\gamma}) \, ,
\label{eqn:gamma_theta}
\end{align}
where $\theta_{b}$, $\theta_{d}$, and $\theta_{\gamma}$ are the
divergence of the baryon, dark-matter, and photon fluid
velocities respectively and an overdot represents a derivative
with respect to the conformal time $\eta$. Here $h$ is the trace
of the spatial metric perturbations $h_{ij}$.  Its evolution is
described by the linearized Einstein equations, which close this
system of linearized equations.  The last terms on the
right-hand-sides of Eqs.~(\ref{eqn:baryon_theta}) and
(\ref{eqn:gamma_theta}) account for Thomson scattering between
baryons and photons, and are responsible for keeping them
tightly coupled in the early Universe.  In these equations
$\sigma_{T}$ is the Thomson cross section, $n_{e}$ is the
electron number density, and $c_{s}$ is the intrinsic sound
speed of the baryons. During tight coupling the second moment
$\Theta_{\gamma}$ of the photon distribution and other higher moments can
be neglected, and the radiation can reliably be given the fluid
description described above.

\begin{figure*}
\centerline{\psfig{file=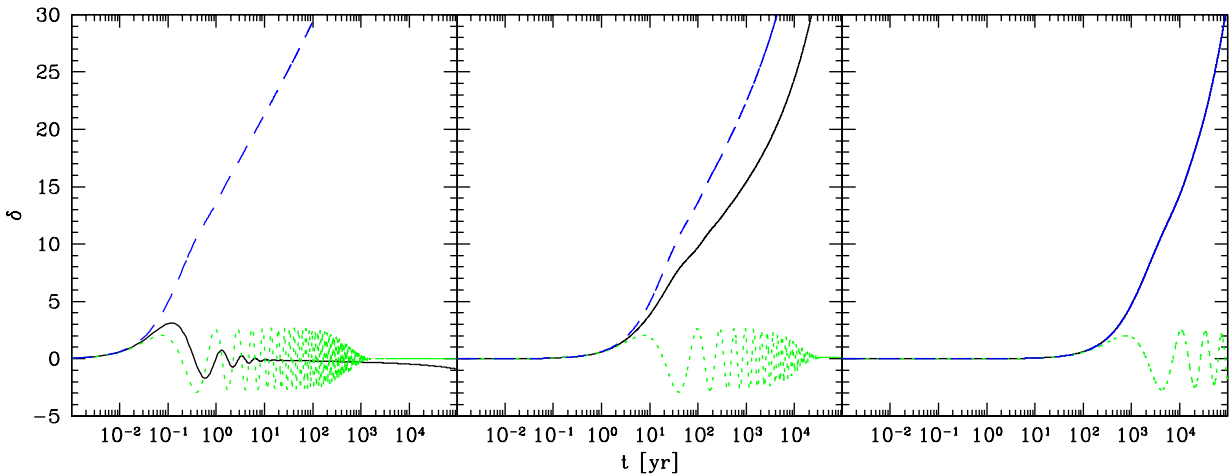,height=2.7in,angle=0}}
\caption{
The evolution of the comoving wavenumber $k=30.0~{\rm Mpc}^{-1}$ (left panel), $k=3.0~{\rm Mpc}^{-1}$ (center panel), and $k=0.3~{\rm Mpc}^{-1}$ density perturbations in the early Universe for dark matter in the $\Lambda$CDM model  (dashed line) and in the model with $\tau=3.5~{\rm yr}$ (solid line).  The `$\beta$' perturbation is represented by the dotted line.  Due to being sourced by the low amplitude `$\beta$' perturbations at early times the dark matter perturbation in the model with a decaying charged component is suppressed relative to the standard $\Lambda$CDM case for $k = 3.0~{\rm Mpc}^{-1}$. For $k \gg 3.0~{\rm Mpc}^{-1}$ (very small scales) $\delta_{d}$ tracks the oscillations in $\delta_{\beta}$ before decay, while for $k \ll 3.0~{\rm Mpc}^{-1}$ (large scales) $\delta_{d}$ follows the standard growing evolution.}
\label{fig:delta_tau}
\end{figure*}

We now show how Eqs.~(\ref{eqn:dark_delta})--(\ref{eqn:gamma_theta}) are modified by the decay of a long-lived metastable charged particle to dark matter in the early Universe. We assume that the decay is of the form $q^{\pm} \rightarrow \ell^{\pm}d$, so the decay of each charged particle $q^{\pm}$ produces a dark-matter particle $d$ and a charged lepton $\ell^{\pm}$.  Denoting the decaying charged component by the subscript `$q$', the background density $\rho_{q}$ evolves according to the equation,
\begin{align}
\dot{\rho}_{q} = -3\frac{\dot{a}}{a}\rho_{q} - \frac{a}{\tau}\rho_{q} \, ,
\label{eqn:background_q}
\end{align}
where $\tau$ is the lifetime of $q^{\pm}$.  The first term just accounts for the normal $a^{-3}$ scaling of non-relativistic matter in an expanding universe, while the second leads to the expected exponential decay of the comoving density.  For the dark matter we have
\begin{align}
\dot{\rho}_{d} = -3\frac{\dot{a}}{a}\rho_{d} + \lambda \frac{a}{\tau}\rho_{q} \, ,
\label{eqn:background_d}
\end{align}
where $\lambda=m_{d}/m_{q}$ is the ratio of the masses of dark matter particle the charged particle. The energy density in photons evolves according to
\begin{align}
\dot{\rho}_{\gamma} = -4\frac{\dot{a}}{a}\rho_{\gamma} + (1-\lambda)\frac{a}{\tau}\rho_{q} \, .
\label{eqn:background_gamma}
\end{align}
This last equation follows from the assumption that the produced lepton
initiates an electromagnetic cascade which rapidly (compared to
the expansion timescale) thermalizes with the photon
distribution.  In practice the last term on the right-hand-side
of Eq.~(\ref{eqn:background_gamma}) is negligibly small because
the decay takes place during the radiation dominated era when
$\rho_{\gamma} \gg \rho_{q}$.  Furthermore,  limits on the
magnitude of $\mu$-distortions to the blackbody spectrum of the
CMB constrain $|1-\lambda|$ to be a small number, as we discuss
below.

Using covariant generalizations of
Eqs.~(\ref{eqn:background_q})--(\ref{eqn:background_gamma}) we
can derive how
Eqs.~(\ref{eqn:dark_delta})--(\ref{eqn:gamma_theta}) are
modified by the transfer of energy and momentum from the `$q$'
component to the dark matter during the decay process.  Since
the charged `$q$' component and the baryons are tightly coupled
via Coulomb scattering they share a common velocity
$\theta_{\beta} = \theta_{b}=\theta_{q}$.
This makes it useful to describe them in terms of a total
charged-species component with energy density
$\rho_{\beta}=\rho_{b}+\rho_{q}$, which we denote here by the
subscript `$\beta$'. Because in the synchronous gauge the decay
proceeds everywhere at the same rate this description is even
more useful as $\delta_{\beta}=\delta_{b}=\delta_{q}$ is
maintained at all times for adiabatic initial conditions.  In
terms of these `$\beta$' variables, then, we have

\begin{align}
\dot{\delta}_{d}  = - \theta_{d}-\frac{1}{2}\dot{h} + \lambda \frac{\rho_{q}}{\rho_{d}}\frac{a}{\tau}(\delta_{\beta}-\delta_{d}) \, ,
\label{eqn:deltadot_d_2}
\end{align}
\begin{align}
\dot{\theta}_{d} = -\frac{\dot{a}}{a}\theta_{d} + \lambda \frac{\rho_{q}}{\rho_{d}}\frac{a}{\tau}(\theta_{\beta}-\theta_{d}) \, ,
\label{eqn:thetadot_d_2}
\end{align}
\begin{align}
\dot{\delta}_{\beta}  = - \theta_{\beta}-\frac{1}{2}\dot{h} \, ,
\label{eqn:delta_beta_2}
\end{align}
\begin{align}
\dot{\theta}_{\beta} = -\frac{\dot{a}}{a}\theta_{\beta} &+ c_{s}^2k^2\delta_{\beta} + \frac{4\rho_{\gamma}}{3\rho_{\beta}} a n_{e} \sigma_{T} (\theta_{\gamma}-\theta_{\beta}) \, ,
\label{eqn:theta_beta_2}
\end{align}
\begin{align}
\dot{\delta}_{\gamma} = -\frac{4}{3}\theta_{\gamma}-\frac{2}{3}\dot{h} + (1-\lambda) \frac{\rho_{q}}{\rho_{\gamma}}\frac{a}{\tau}(\delta_{\beta}-\delta_{\gamma})\, ,
\label{eqn:deltadot_gamma_2}
\end{align}
and
\begin{align}
\dot{\theta}_{\gamma} = k^2 \left( \frac{1}{4}\delta_{\gamma} - \Theta_{\gamma} \right) &+ a n_{e}\sigma_{T}(\theta_{\beta}-\theta_{\gamma}) \nonumber \\
&+ (1-\lambda) \frac{\rho_{q}}{\rho_{\gamma}}\frac{a}{\tau}\left(\frac{3}{4}\theta_{\beta}-\theta_{\gamma}\right) \, .
\label{eqn:thetadot_gamma_2}
\end{align}

We now describe how small-scale modes that enter the horizon
prior to decay are suppressed relative to those modes that enter
the horizon after decay.  Due to the Thomson collision terms the
`$\beta$' component and the photons will be tightly coupled as a
`$\beta$'-photon fluid at early times and this fluid will
support acoustic oscillations.  Furthermore,
Eqs.~(\ref{eqn:deltadot_d_2}) and (\ref{eqn:thetadot_d_2}) show
that the dark-matter perturbations are strongly sourced by the
perturbations of the `$\beta$' component prior to decay, when
the ratio $\rho_{q}/\rho_{d}$ is large.  Dark-matter modes that
enter the horizon prior to decay will thus track the
oscillations of the `$\beta$'-photon fluid rather than simply
growing under the influence of gravity.  After decay, when the
ratio $\rho_{q}/\rho_{d}$ is small, the source term shuts off
and dark-matter modes that enter the horizon undergo the
standard growing evolution. In Fig.~\ref{fig:delta_tau} we
follow the evolution of the dark-matter perturbations through
the epoch of decay.  We modified {\tt CMBFAST} \cite{CMBfast} to
carry out these calculations.

\begin{figure}
\centerline{\psfig{file=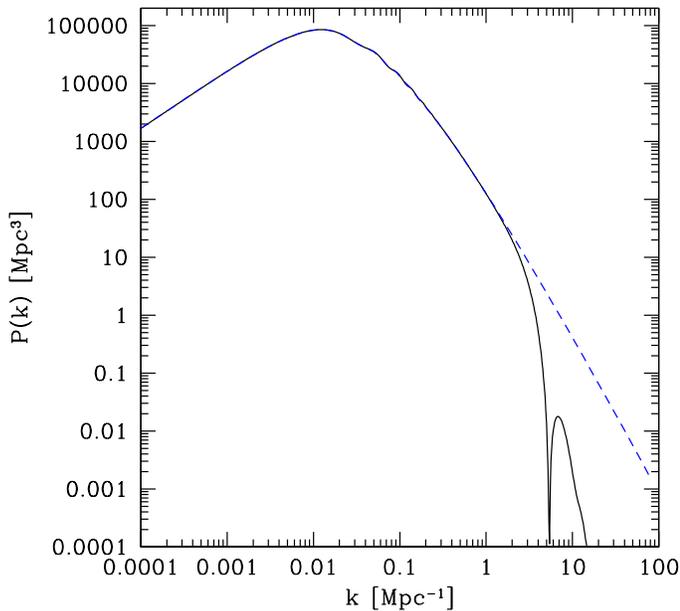,width=3.6285in,angle=0}}
\caption{The linear-theory power spectrum of matter density fluctuations in the standard $\Lambda$CDM model (dashed line), and in the charged decay to dark matter model (solid line) with $\tau = 3.5~{\rm yr}$.  The charged decay model matches the standard $\Lambda$CDM model on length scales larger than $0.3~{\rm Mpc}$, but power drops of sharply below $0.3~{\rm Mpc}$.}
\label{fig:pow}
\end{figure}

In order to suppress power on subgalactic scales the decay
lifetime must be roughly the age of the Universe when the mass
enclosed in the Hubble volume is equal to a galaxy mass; this
occurs when $\tau \sim$ years. In Fig.~\ref{fig:pow} we plot the
linear power spectrum of matter density fluctuations at the
present day for a charged-particle lifetime $\tau  = 3.5~{\rm
yr}$ assuming a scale-invariant primordial power spectrum. We
see that power is suppressed on scales smaller than $k^{-1} \sim
0.3~ {\rm Mpc}$ relative to the standard $\Lambda$CDM power
spectrum.  Suppression of power on these length scales 
reduces the expected number of
subgalactic halos, bringing the predictions in line with
observation \cite{Kamion2000} without violating constraints from
the Lyman-alpha forest \cite{White2000}.  Of course, the model
reproduces the successes of the standard $\Lambda$CDM model on
larger scales and in the CMB.

The requirements of the charged-particle species are that it have a comoving mass density equal to the dark-matter density today and have a lifetime of $\tau \sim$ 3.5 yr.  In order to satisfy the constraint to the CMB chemical potential \cite{Fixsen1996}, the fractional mass difference between the charged and neutral particles must be $\Delta m/m < 3.6 \times 10^{-3}$, and in order for the decay to be allowed kinematically the mass difference must be greater than the electron mass.  One possibility is the SuperWIMP scenario of Ref.~\cite{Feng2003} in which a charged particle may decay to an exclusively gravitationally interacting particle.
For example, in supersymmetric models, the decay of a selectron to an electron and gravitino $\widetilde{e} \rightarrow e\,\widetilde{G}$ with $m_{\widetilde{e}} \approx m_{\widetilde{G}} > 122~{\rm TeV}$ would satisfy these constraints, as would the decay of a KK-electron to an electron and KK-graviton $e^{1} \rightarrow e\,G^{1}$ with $m_{e^{1}} \approx m_{G^{1}} > 72~{\rm TeV}$  in the case of the single universal extra dimension Kaluza-Klein (KK) model discussed in Refs.~\cite{Appel2001,Feng2003}.  Such masses are larger than the unitary bound for thermal production \cite{Griest1990}, but might be accommodated through nonthermal mechanisms or if the next-to-lightest partner is a squark which might then interact more strongly and thus evade this bound.  There may also be viable scenarios involving nearly-degenerate charged and neutral higgsinos.

It should be noted that the recent WMAP evidence for early star
formation \cite{Kogut2003} argues against the suppression of
small-scale power, but these results are not yet conclusive.  If
it does turn out that traditional astrophysical mechanisms can
explain the dearth of dwarf galaxies, then our arguments can be
turned around to provide constraints to an otherwise inaccessible
region of the parameter space for decaying dark matter
\cite{SigKam}.  Finally, if the mechanism we propose here is
realized in nature, then the dearth of small-scale power, along
with the detection of a non-zero CMB chemical potential, would
be a powerful probe of the particle spectrum of the new physics
responsible for dark matter.

\begin{acknowledgments}
KS acknowledges the support of a
Canadian NSERC Postgraduate Scholarship.  This work was
supported in part by NASA NAG5-9821 and DoE DE-FG03-92-ER40701.
\end{acknowledgments}

\end{document}